\documentclass[12pt]{article}

\usepackage[cp1251]{inputenc}
\usepackage[english,russian]{babel}
\usepackage{amsfonts}
\usepackage{amscd}
\usepackage{amsmath}
\usepackage{amssymb}
\usepackage{latexsym}
\usepackage{longtable}

\begin{document}

\begin{center}
\large  Non-existence of a short algorithm for multiplication \\ 
 of $3\times3$ matrices with group $S_4\times S_3$, II \\
\medskip
\normalsize Vladimir~P.~Burichenko  \\
\medskip
\small Institute of Mathematics of the National Academy of Sciences of Belarus \\
 e-mail: vpburich@gmail.com 
\end{center}
\bigskip

\begin{abstract}\noindent
It is proved that there is no an algorithm for multiplication of
$3\times3$ matrices of multiplicative length $\leq23$ that is
invariant under a certain group isomorphic to $S_4\times S_3$. The
proof makes use of description of the orbits of this group on
decomposable tensors in the tensor cube $(M_3({\mathbb
C}))^{\otimes3}$ which was obtained earlier.

({\em MSC classification 68Q25, 20C}).
\end{abstract}

{\bf 1. Introduction.}\ \  The present work is concerned with the
problem of fast matrix multiplication, namely studying of
algorithms with a nontrivial symmetry group. We show that there
exists no an algorithm for multiplication of $3\times3$ matrices
of length $\leq23$ that is invariant under a certain group $G$
isomorphic to $S_4\times S_3$. This article is an immediate sequel
of article~[1]. The more detailed discussion, motivation, and
further references can be found in~[1]. Here we restrict ourselves
with stating Theorem 1 of [1] (whose proof is the main aim of the
present work), as well as the main result of [1], namely the
classification of orbits on the decomposable tensors.

For convenience of the reader who is not very experienced in algorithms
we now state the result we are going to prove in purely group- and
representation-theoretic terms. Let
$$ M=M_3({\mathbb C})=\langle e_{ij}\mid 1\leq i,j\leq3\rangle_{\mathbb C}$$
be the space of complex $3\times3$ matrices. Consider the tensor
$$ {\mathcal T}=\sum_{1\leq i,j,k\leq3}e_{ij}\otimes  e_{jk}\otimes
  e_{ki} \in M\otimes M\otimes M\,. $$

Let $A\leq GL(3,{\mathbb C})$ be the group of all monomial $3\times3$
matrices
whose nonzero elements are $\pm1$ and the determinant is $\det=1$. It
is easy to see that $A\cong S_4$, and $A$ is irreducible. This group
$A$ acts on $M^{\otimes3}$ ``diagonally'', that is, $a\in A$ acts by a
transformation
$$ T(a)\colon x\otimes y\otimes z\mapsto axa^{-1}\otimes aya^{-1}
  \otimes aza^{-1}\,.$$
It may be shown that this action of $A$ preserves ${\mathcal T}$.

Next, consider the following transformations:
$$ \rho(x\otimes y\otimes z)=y^t\otimes x^t\otimes z^t\,,\qquad
  \sigma(x\otimes y\otimes z)  =z\otimes x\otimes y\,$$
(where $t$ means transpose). It is easy to see that both $\rho$ and
$\sigma$ preserve ${\mathcal T}$, and $B:=\langle\rho,\sigma\rangle
 \cong S_3$.
Finally, it is not hard to show that $A$ and $B$ commute elementwise
(for the details of these (and even more general) calculations
the reader can consult [2] or~[3]). Thus, the group $G=A\times B\cong
S_4\times S_3$ acts on $M^{\otimes3}$ and preserves ${\mathcal T}$.

The tensors of $M^{\otimes3}$ of the form $v_1\otimes v_2\otimes v_3$
will be called
{\em elementary}, or {\em decomposable.} A {\em decomposition} of
length $l$ for ${\mathcal T}$ is an (unordered) set of $l$
elementary tensors
$$ {\mathcal P}=\{t_i=x_i\otimes y_i\otimes z_i\mid i=1,\ldots,l\}$$
such that $t_1+\ldots+t_l={\mathcal T}$.

Obviously, any element of $G$ takes a length $l$ decomposition to a
length $l$ decomposition. In particular, we can consider a notion of
a $G$-invariant decomposition. Now we can state the main result of
the present article.
\smallskip

{\bf Theorem 1.} {\em Let ${\mathcal T}$ and $G=A\times B$ be as described
above. Then there exists no $G$-invariant decomposition of ${\mathcal T}$ of
length $\leq23$. }
\smallskip

To prove this theorem it is necessary, first of all, to describe
all orbits of length $\leq23$ for the group $G$ on the
decomposable tensors in $M^{\otimes3}$. This is done in [1] (in
fact, it is sufficient to consider orbits of length $\leq18$,
because $|G|$ is not divisible by $19\leq d\leq23$ ).

Below in the article ${\rm St}_G(w)$ is the stabilizer of a tensor
$w$ with respect to the action of $G$; $\zeta$ is the primitive
cubic root of $1$, and $i=\sqrt{-1}$ (we use the same symbol $i$
for indices, but hope that this will not lead to a confusion even
in the formulae like $e_{ij}-ie_{ki}$). Also,
$$ \delta=e_{11}+e_{22}+e_{33}\,,\quad \varkappa=\sum_{i\ne j}e_{ij}
  =e_{12}+e_{21}+e_{13}+e_{31}+e_{23}+e_{32}\,,$$
$$\eta=e_{11}+\zeta e_{22}+\overline\zeta e_{33}\,,\qquad \overline
  \eta=  e_{11}+\overline\zeta e_{22}+   \zeta e_{33}\,,$$
$$\tau=e_{12}+e_{23}+e_{31}-e_{21}-e_{32}-e_{13}\,. $$

In [1] the following was proved.

{\bf Proposition 2.} {\em Any orbit of length $\leq18$ of $G$ on
decomposable tensors in $M^{\otimes3}$ has a representative of the form
$w_i(a,b,\ldots)$, $1\leq i\leq44$, where $w_i$ are the tensors listed
in the following table.

\begin{longtable}{|p{0.5cm}|p{0.5cm}|p{12cm}|}
\hline
\rule{0pt}{3ex}  $i$ & $l_i$ & $w_i(a,b,\ldots)$ \\
\hline  \endhead
  $1$ &  $12$ & \rule[-2ex]{0pt}{5ex}
  $(a(e_{11}+e_{22})+b(e_{12}+e_{21})+ce_{33}+d(e_{13}+e_{23}+e_{31}
  +e_{32}))^{\otimes3}$ \\
\hline  $2$ &  $12$ & \rule[-2ex]{0pt}{5ex}
  $ (ae_{11}+be_{22}+ce_{33}+d(e_{12}+e_{21}))^{\otimes3}$ \\
\hline $3$   &  $6$ & \rule[-2ex]{0pt}{5ex}
   $ (a(e_{11}+e_{22})+be_{33}+c(e_{12}+e_{21}))^{\otimes3}$ \\
\hline $4$  &  $6$ & \rule[-2ex]{0pt}{5ex}
   $(ae_{11}+be_{22}+ce_{33})^{\otimes3}$ \\
\hline $5$ &  $3$ & \rule[-2ex]{0pt}{5ex} $(a(e_{11}+e_{22})+be_{33}
 )^{\otimes3}$ \\
\hline $6$ &  $2$ & \rule[-2ex]{0pt}{5ex}
   $a\eta^{\otimes3}$ \\
\hline $7$ &  $1$ & \rule[-2ex]{0pt}{5ex} $a\delta^{\otimes3}$ \\
\hline $8$  &  $16$ & \rule[-2ex]{0pt}{5ex}
	$(a\eta+b(e_{12}+\zeta e_{23}+\overline\zeta e_{31})
	+c(e_{21}+\zeta e_{32}+\overline\zeta e_{13}))^{\otimes3}$ \\
\hline $9$ & $4$ & \rule[-2ex]{0pt}{5ex} $(a\delta+b\varkappa)^
  {\otimes3}$ \\
\hline $10$  &  $8$ & \rule[-2ex]{0pt}{5ex}
	$(a\eta+b(e_{12}+e_{21}+\zeta(e_{23}+e_{32})+
	  \overline\zeta(e_{31}+e_{13})))^{\otimes3}$ \\
\hline $11$  &  $8$ & \rule[-2ex]{0pt}{5ex}
	$(a\delta+b(e_{12}+e_{23}+e_{31})+c(e_{21}+e_{32}+e_{13}))
  ^{\otimes3}$ \\
\hline $12$ &  $6$ & \rule[-2ex]{0pt}{5ex}
   $(a(e_{11}+e_{22})+b(e_{12}-e_{21})+ce_{33})^{\otimes3}$ \\
\hline $13$ &  $12$ & \rule[-2ex]{0pt}{5ex}
   $(a(e_{11}+e_{22})+b(e_{12}+e_{21})+ce_{33}+d(e_{13}+e_{23}
  -e_{31}-e_{32}))^{\otimes3}$ \\
\hline $14$ &  $12$ & \rule[-2ex]{0pt}{5ex}
   $(a(e_{11}+e_{22})+be_{12}+ce_{21}+de_{33})^{\otimes3}$ \\
\hline $15$ &  $12$ & \rule[-2ex]{0pt}{5ex}
   $(ae_{11}+be_{22}+c(e_{12}-e_{21})+de_{33})^{\otimes3}$ \\
\hline  $16$ & $18$ &  \rule[-2ex]{0pt}{5ex}
$(a(e_{11}+e_{22})+be_{33})\otimes
(c(e_{11}+e_{22})+de_{33})\otimes(f(e_{11}+e_{22})+ge_{33})$     \\
\hline $17$ &  $18$ & \rule[-2ex]{0pt}{5ex} $a(e_{11}-e_{22})\otimes
  (e_{12}+e_{21})\otimes(e_{12}-e_{21})$  \\
\hline $18$ &  $9$ & \rule[-2ex]{0pt}{5ex} $(e_{11}-e_{22})^{\otimes2}
\otimes(a(e_{11}+e_{22})+be_{33})$  \\
\hline $19$ &  $9$ & \rule[-2ex]{0pt}{5ex} $(e_{12}+e_{21})^{\otimes2}
\otimes(a(e_{11}+e_{22})+be_{33})$   \\
\hline $20$  &  $9$ & \rule[-2ex]{0pt}{5ex} $(e_{12}-e_{21})^{\otimes2}
\otimes(a(e_{11}+e_{22})+be_{33})$  \\
\hline $21$  &  $9$ & \rule[-2ex]{0pt}{5ex}
$(a(e_{11}+e_{22})+be_{33})^{\otimes2}\otimes(c(e_{11}+e_{22})
  +de_{33})$   \\
\hline $22$   &  $18$ & \rule[-2ex]{0pt}{5ex} $(a(e_{11}+e_{22})+
b(e_{12}+e_{21})+ce_{33})^{\otimes2}\otimes(d(e_{11}+e_{22})+
  f(e_{12}+e_{21})+ge_{33})$   \\
\hline $23$   &  $18$ & \rule[-2ex]{0pt}{5ex}
$(a(e_{11}-e_{22})+b(e_{12}-e_{21}))\otimes (a(e_{11}-e_{22})-b(e_{12}
  -e_{21}))\otimes (c(e_{11}+e_{22})+d(e_{12}+e_{21})+fe_{33})$  \\
\hline $24$   &  $18$ & \rule[-2ex]{0pt}{5ex}
$(a(e_{13}+e_{23})+b(e_{31}+e_{32}))\otimes (b(e_{13}+e_{23})+
  a(e_{31}+e_{32}))\otimes(c(e_{11}+e_{22})+d(e_{12}+e_{21})+fe_{33})$  \\
\hline $25$   &  $18$ & \rule[-2ex]{0pt}{5ex}
$(ae_{11}+be_{22}+ce_{33})^{\otimes2}\otimes (de_{11}+fe_{22}+ge_{33})$ \\
\hline $26$   &  $18$ & \rule[-2ex]{0pt}{5ex}
$(ae_{12}+be_{21})\otimes(be_{12}+ae_{21})\otimes(ce_{11}+de_{22}
  +fe_{33})$  \\
\hline $27$   &  $18$ & \rule[-2ex]{0pt}{5ex}
$ (a(e_{11}+e_{22})+b(e_{12}-e_{21})+ce_{33})\otimes (a(e_{11}+
  e_{22})-b(e_{12}-e_{21})
+ce_{33})\otimes (d(e_{11}+e_{22})+fe_{33})$ \\
\hline $28$   &  $18$ & \rule[-2ex]{0pt}{5ex}
$(a(e_{11}-e_{22})+b(e_{12}+e_{21}))^{\otimes2}\otimes(c(e_{11}+
  e_{22})+de_{33})$ \\
\hline $29$   &  $18$ & \rule[-2ex]{0pt}{5ex}
$(a(e_{13}+ie_{23})+b(e_{31}+ie_{32}))\otimes (b(e_{13}+ie_{23})+
  a(e_{31}+ie_{32}))\otimes(c(e_{11}-e_{22})+d(e_{12}+e_{21}))$ \\
\hline $30$   &  $18$ & \rule[-2ex]{0pt}{5ex}
$(a(e_{11}+e_{22})+b(e_{12}+e_{21})+ce_{33})\otimes(a(e_{11}+e_{22})
 -b(e_{12}+e_{21})+ce_{33})\otimes(d(e_{11}+e_{22})+fe_{33})$ \\
\hline $31$   &  $18$ & \rule[-2ex]{0pt}{5ex}
$(a(e_{11}-e_{22})+b(e_{12}-e_{21}))^{\otimes2}\otimes(c(e_{11}+
  e_{22})+de_{33})$ \\
\hline $32$   &  $18$ & \rule[-2ex]{0pt}{5ex}
$(a(e_{13}+e_{23})+b(e_{31}+e_{32}))\otimes (b(e_{13}-e_{23})+
  a(e_{31}-e_{32}))\otimes(c(e_{11}-e_{22})+d(e_{12}-e_{21}))$ \\
\hline $33$   &  $18$ & \rule[-2ex]{0pt}{5ex}
$(ae_{11}+be_{22}+ce_{33})\otimes (be_{11}+ae_{22}+ce_{33})\otimes
  (d(e_{11}+e_{22})+fe_{33})$ \\
\hline $34$   &  $18$ & \rule[-2ex]{0pt}{5ex}
$(ae_{12}+be_{21})^{\otimes2}\otimes(c(e_{11}+e_{22})+de_{33})$ \\
\hline $35$   &  $18$ & \rule[-2ex]{0pt}{5ex}
$(ae_{13}+be_{31})\otimes(be_{23}+ae_{32})\otimes(ce_{12}+de_{21})$ \\
\hline $36$   &  $18$ & \rule[-2ex]{0pt}{5ex}
$(a(e_{11}+e_{22})+b(e_{12}-e_{21})+ce_{33})^{\otimes2}\otimes
  (d(e_{11}+e_{22})+f(e_{12}-e_{21})+ge_{33})$ \\
\hline $37$   &  $18$ & \rule[-2ex]{0pt}{5ex}
$(a(e_{11}-e_{22})+b(e_{12}+e_{21}))\otimes(a(e_{11}-e_{22})-b(e_{12}
  +e_{21}))\otimes(c(e_{11}+e_{22})+d(e_{12}-e_{21})+fe_{33})$ \\
\hline $38$   &  $18$ & \rule[-2ex]{0pt}{5ex}
$(a(e_{13}+ie_{23})+b(e_{31}+ie_{32}))\otimes (b(e_{13}-ie_{23})+
 a(e_{31}-ie_{32}))\otimes(c(e_{11}+e_{22})+d(e_{12}-e_{21})+fe_{33})$ \\
\hline $39$ &  $6$ & \rule[-2ex]{0pt}{5ex} $a\eta\otimes\overline\eta
  \otimes\delta$  \\
\hline  $40$ &  $12$ & \rule[-2ex]{0pt}{5ex}
$(a\delta+b\varkappa)^{\otimes2}\otimes(c\delta+d\varkappa)$ \\
\hline  $41$ &  $12$ & \rule[-2ex]{0pt}{5ex} $\tau^{\otimes2}\otimes
(a\delta+b\varkappa)$ \\
\hline  $42$ &   $12$ & \rule[-2ex]{0pt}{5ex} $(ae_{11}+be_{22}
  +ce_{33})\otimes(ce_{11}+ae_{22}+be_{33})\otimes(be_{11}+ce_{22}
  +ae_{33})$  \\
\hline  $43$ &  $6$ & \rule[-2ex]{0pt}{5ex} $(ae_{11}+b(e_{22}+e_{33}))
 \otimes(ae_{22}+b(e_{11}+e_{33}))\otimes (ae_{33}+b(e_{11}+e_{22}))$
 \\
\hline  $44$ &  $6$ & \rule[-2ex]{0pt}{5ex}
$(ae_{23}+be_{32})\otimes(be_{13}+ae_{31})\otimes(ae_{12}+be_{21})$ \\
\hline
\end{longtable}

}

This proposition is the Theorem 4 of [1], slightly shortened. Here
$l_i$ is the length of the orbit. The number $i$ (the number of the
row) will be referred to as the {\em type} of the tensor $w_i(a,b,
\ldots)$ (and of its orbit).

It should be noted that

1) The parameters $a,b,\ldots$ for the tensor $w_i(a,b,\ldots)$,
which is a representative of a given orbit, are not uniquely
defined, in general. Particularly, in most part of cases we have
$w_i(a,b,\ldots) =w_i(\zeta^l a,\zeta^lb,\ldots)$, where
$l=0,1,2$. Moreover, there are other situations, where the orbits
of two tensors $w_i(a,b,\ldots)$ and $w_i(a',b',\ldots)$ coincide,
but $(a,b,\ldots)\ne(a',b',\ldots)$ (see [1] for details).

2) For some ``degenerate'' $a,b,\ldots$ the length of the orbit of
$w_i(a,b,\ldots)$ can be less than $l_i$ (in fact, this length is
the proper divisor of $l_i$). In such a case there exists a type
$j\ne i$ an some parameters $a',b',\ldots$ such that $w_i(a,b,\ldots)=
w_j(a',b',\ldots)$, and $(a',b',\ldots)$ is nondegenerate for type $j$.

For instance, let $i=4$,
$w_4(a,b,c)=(ae_{11}+be_{22}+ce_{33})^{\otimes3}$. Then the orbit
of $w_4(a,b,c)$ has $6$ points when $a$, $b$, and $c$ are pairwise
distinct. If there are exactly $2$ distinct among them, then the
orbit has length $3$ and is generated by a tensor of the form
$w_5(a',b')$. Say, if $a\ne b=c$, then $Gw_4(a,b,c)=Gw_5(a,b)$
(and when $a=b=c$, we have
$w_4(a,a,a)=a^3\delta^{\otimes3}=w_7(a^3)$).

Below $s_i$ is the number of the parameters $a,b,\ldots$ in the tensor
$w_i(a,b,\ldots)$. Also, for each type $i$ let $H_i\leq G$ be the
``typical'' stabilizer of $w_i(a,b,\ldots)$, that is, the stabilizer
for nondegenerate $(a,b,\ldots)$. Say, for $i=4$ the stabilizer $H_4$
is a certain subgroup isomorphic to $Z_2^2\times S_3$, specifically
the subgroup of all elements of the form $(c,b)$, where $b\in B$, and
$c={\rm diag}(\varepsilon_1,\varepsilon_2,\varepsilon_3)\in A$, where
$\varepsilon_i=\pm1$, $\varepsilon_1\varepsilon_2\varepsilon_3=1$.
Clearly, the index $|G:H_i|$ is equal to $l_i$.
\medskip

{\bf 2. Reduction to polynomial systems.} The aim of this section is
to show that the proof of Theorem 1 can be reduced to solution of
several systems of polynomial equations (or, to be more precise, to
the proof that these systems have no solutions).

If $\widetilde V=V_1\otimes\ldots\otimes V_l$ is the tensor product
of several spaces
and $w\in\widetilde V$ is an arbitrary tensor, then finding of all
representations of $w$ as a sum of $\leq r$ decomposable tensors
reduces, as one can easily see, to the solution of a certain system
of polynomial equations (which are known as (generalized) {\em Brent
equations}, after the work [4]). Specifically, let $d_i=\dim V_i$,
$\{v_{ij}\mid 1\leq j\leq d_i\}$ be the bases of $V_i$, and
$w_{k_1\ldots k_l}$ be the coordinates of $w$ in the natural tensor
basis, i.e.,
$$ w=\sum_{1\leq k_i\leq d_i} w_{k_1\ldots k_l} v_{1,k_1}\otimes\ldots
  \otimes v_{l,k_l}\,.$$
Then, clearly, finding all decompositions of $w$ of length $\leq r$
is equivalent to solving the system of $d_1\ldots d_l$ equations
$$ \sum_{j=1}^r x_{1,k_1}^{(j)}\ldots x_{l,k_l}^{(j)}=w_{k_1\ldots k_l}
  \,, \qquad   1\leq k_i\leq d_i $$
in $r(d_1+\ldots+d_l)$ unknowns $x_{i,k_i}^{(j)}$, $1\leq j\leq r$,
$1\leq k_i\leq d_i$.

The latter statement has a ``group-invariant'' version. Namely, if
$X$ is a finite group of linear transformations of $\widetilde V$,
preserving representation of $\widetilde V$ as a tensor product (but
possibly permuting
the factors $V_i$), and $w$ is an $X$-invariant tensor, then finding
all $X$-invariant decompositions of $w$, whose length is $\leq r$,
can be reduced to the solution of some set of polynomial systems. It
is not difficult to prove this statement in the general situation, but
in the present article we restrict ourselves with the particular case
of $\widetilde V=M^{\otimes3}$, $X=G=A\times B$, $w={\mathcal T}$,
and $r=23$.

Let ${\mathcal P}=\{t_i=x_i\otimes y_i\otimes z_i\mid 1\leq i\leq l\}$
be a $G$-invariant decomposition of length $l$ for ${\mathcal T}$. We
have a partition of ${\mathcal P}$ into $G$-orbits: ${\mathcal P}=
{\mathcal O}_1\sqcup\ldots\sqcup{\mathcal O}_q$.
The {\em type} of ${\mathcal P}$ is the multiset $\{n_1,\ldots,n_q\}$,
where $n_i$ is the type of ${\mathcal O}_i$. Clearly, we can assume
that $n_i$ are ordered: $n_1\leq\ldots\leq n_q$. It is also clear that
the length of a decomposition of type $\{n_1,\ldots,n_q\}$ is equal
to $\sum_{i=1}^q l_{n_i}$.

To describe all $G$-invariant decompositions of length $\leq23$ it is
sufficient to describe all $G$-invariant decompositions of a given
type $\{n_1,\ldots,n_q\}$, for every type such that
$\sum_{i=1}^q l_{n_i}\leq23$. Obviously, there exist finitely many
such types. So, to show that the description of all $G$-invariant
decompositions of length $\leq23$ reduces to the solution of some
finitely many polynomial systems, it is sufficient to show that the
description of all $G$-invariant decompositions of a given type
$\{n_1,\ldots,n_q\}$ reduces to solution of several (in fact, one!)
polynomial systems.

Take some representatives $h_{ij}$, $1\leq j\leq l_i$, for cosets
$G/H_i$. Then any orbit of type $i$ is, clearly,
$\{h_{ij}w_i(a_1,\ldots,a_{s_i})\mid j=1,\ldots,l_i\}$ for some
$a_1,\ldots,a_{s_i}\in{\mathbb C}$. So a decomposition of type
$\{n_1,\ldots,n_q\}$ is
$${\mathcal P}=\{ h_{n_i,j}w_{n_i}(a_{i,1},\ldots,a_{i,u_i}) \mid
  1\leq i\leq q,\ 1\leq j\leq l_{n_i}\},$$
where $u_i=s_{n_i}$, for some array $(a_{im}\in{\mathbb C}\mid 1\leq i
 \leq q,\ 1\leq m\leq u_i)$.

The condition that the sum of elements of ${\mathcal P}$ equals
${\mathcal T}$ now takes the following (rather clumsy) form:
$$ \sum_{i=1}^q \sum_{j=1}^{l_{n_i}} h_{n_i,j}w_{n_i}(a_{i,1},\ldots,
  a_{i,u_i})\ =\ {\mathcal T}\,. \eqno(1)$$

The tensor $w_m(a_1,\ldots,a_{s_m})$ depends polynomially of its
parameters, by Proposition~2. So the left-hand side of the latter
condition depends on the parameters $a_{ij}$ polynomially also,
and so equality (1) is equivalent to some system of polynomial
equations in $a_{ij}$, as required.

There exists another condition, which is equivalent to (1), but looks
simpler and does not involve subgroups or cosets.  Note that since
$G$ is finite and the characteristics is $0$, $N=M^{\otimes3}$
decomposes as $N=N^G\oplus N_0$, where $N^G=\{x\in N\mid  gx=x\
\forall\ g\in G\}$
is the subspace of invariants of $G$ in $N$, and $N_0$ is the subspace
of all elements whose averaging over $G$ is $0$:
$$ N_0=\{ x\in N\mid \frac{1}{|G|}\sum_{g\in G} gx\,=\,0\}. $$
By $p$ we denote averaging operator, i.e., $p(x)=(1/|G|)\sum_{g\in G}gx$.
It is clear that $p$ is nothing else but the projection onto $N^G$
parallel to $N_0$ : $p={\rm pr}_{N^G}$.

Let $H\leq G$ be an arbitrary subgroup of index $l=|G:H|$,
$g_1,\ldots,g_l$ be the representatives of the cosets $G/H$, and
let $w\in N$ be an $H$-invariant tensor (not decomposable, in general).
Then the $G$-orbit of $w$ is $Gw=\{g_iw\mid i=1,\ldots,l\}$.
(Strictly speaking, if we consider $\{g_iw\mid i=1,\ldots,l\}$ as a
multiset, then it is an integer multiple of an orbit, of multiplicity
$|H_1:H|$, where $H_1={\rm St}_G(w)$ is the stabilizer of~$w$. But we
neglect the possibility that $H_1>H$, for simplicity). And it is clear
that the sum of elements of an orbit is $\sum_{i=1}^l g_iw=lp(w)$.
Hence the condition (1) can be restated as
$$\sum_{i=1}^q l_{n_i}p(w_{n_i}(a_{i,1},\ldots,a_{i,u_i}))={\mathcal T}. \eqno(2)$$

{\bf Remark.} Strictly speaking, the condition (1), or equivalently
(2), should be augmented by the requirement that $(a_{i,1},\ldots,
a_{i,u_i}))$ is a nondegenerate array of parameters for type $n_i$.
But if this array of parameters is degenerate, then
$$\{ h_{n_i,j}w_{n_i}(a_{i1},\ldots,a_{i,u_i})
  \mid 1\leq j\leq l_{n_i}\}$$
is an integer multiple (of multiplicity $>1$) of an orbit of smaller
length, and we obtain a $G$-invariant decomposition for ${\mathcal T}$
whose length is $<\sum_{i=1}^q l_{n_i}$. (It should be noticed here
that always $zw_i(a,b,\ldots)=w_i(z'a,z'b,\ldots)$, for any
$z\in{\mathbb C}$, where $z'=z$ for $i=6,7,17,18,19,20,39,41$ and
$z'=z^{1/3}$ for the other $i$.)

This way or that, but we see that the statement that studying of
$G$-invariant decompositions of length $\leq23$ for ${\mathcal T}$
reduces to solution of several polynomial systems, is still true,
despite of possibility of degenerate arrays of parameters.
\medskip

{\bf 3. The subspace of $G$-invariants.}

In this section we consider the subspace $R=N^G$ and the projection
onto $R$ in more details.

Let $F$ be the set of ordered triples of ordered pairs of elements of
$\{1,2,3\}$:
$$ F=\{((i_1,j_1),(i_2,j_2),(i_3,j_3)) \mid i_k,j_k\in\{1,2,3\}\}. $$
That is, $F$ is precisely the set of ``indices'' for the standard
basis of $N$  :
$$ N=\langle e_\alpha\mid \alpha\in F\rangle_{\mathbb C}\,,\qquad
  e_\alpha= e_{i_1j_1}\otimes e_{i_2j_2}\otimes e_{i_3j_3}\,,\quad
  \alpha=((i_1,j_1),  (i_2,j_2),(i_3,j_3)). $$

Note that $F$ is acted on by group $S_3\times S_3$. The first $S_3$
acts on indices:
$$ (g,1)\ ((i_1,j_1),(i_2,j_2),(i_3,j_3))=((gi_1,gj_1),(gi_2,gj_2),
  (gi_3,gj_3))\,,\quad g\in S_3. $$
The second factor permutes the pairs, and transposes each pair, if
the acting element is odd:
$$ (1,(123))\ ((i_1,j_1),(i_2,j_2),(i_3,j_3))=((i_3,j_3),(i_1,j_1),
  (i_2,j_2))\,, $$
$$ (1,(12))\ ((i_1,j_1),(i_2,j_2),(i_3,j_3))=((j_2,i_2),(j_1,i_1),
   (j_3,i_3))\,. $$
It is not difficult to check that with these definitions we obtain
an action of $S_3\times S_3$ indeed; the details are left to the reader.

Consider natural homomorphisms $A\longrightarrow S_3$ and
$B\longrightarrow S_3$. Namely, to a matrix $a\in A$ corresponds the
permutation of the lines $\langle e_1\rangle$, $\langle e_2\rangle$,
$\langle e_3\rangle$ induced by~$a$.
And to an element $b\in B$ corresponds the permutation of factors in
the tensor product $M\otimes M\otimes M$, associated to $b$. Now we
can define a homomorphism $\varphi\colon G=A\times B\longrightarrow
S_3\times S_3$, ``by components''. We denote $\varphi(g)$ also by
$\overline g$.

It is convenient to consider a group slightly larger than $G$,
namely $G_1=A_1\times B$, where $A_1$ is the group of all (that
is, not necessary of determinant $+1$) monomial $3\times3$
matrices whose nonzero elements are $\pm1$. Obviously,
$A_1=A\times \langle-E\rangle_2$, where $E$ is the identity
matrix, whence $G_1=G\times\langle-E\rangle_2$. However, the
action of $G_1$ on $N$ reduces to the action of $G$, because,
clearly, $T(-E)={\rm id}_N$. Also, let $C_1=\{{\rm
diag}(\varepsilon_1,\varepsilon_2,\varepsilon_3)
\mid\varepsilon_i=\pm1\}$, and $C=C_1\cap G$ be the subgroup of
matrices satisfying $\varepsilon_1\varepsilon_2\varepsilon_3=1$.
It is obvious that $C_1=C\times\langle-E\rangle_2$.

The advantage of considering $G_1$ is that all permutation matrices
are in $A_1$, and any element of $A_1$ is
uniquely representable in the form $a=c\widehat\pi$, where $\pi$ is
the permutation, corresponding to $a$, $\widehat\pi$ is the corresponding
permutation matrix, and $c\in C_1$ .

It is easy to note that $G_1$ permutes the elements of the standard
basis $\{e_\alpha\}$ up to  sign, that is, the set
$\{\pm e_\alpha\mid \alpha\in F\}$ is $G_1$-invariant. More precisely,
the following fact is true.
\smallskip

{\bf Lemma 3.} {\em For any $\alpha\in F$ and $g\in G_1$ holds
$ge_\alpha=\pm e_{\overline g\alpha}$. }

{\bf Proof.} This statement is easy, nevertheless we give a detailed
proof. First of all, if the desired equality is true for two elements
$g,h\in G_1$ and for all $\alpha\in F$, then it is true for $gh$ also.
Indeed,
$$ (gh)e_\alpha=g(he_\alpha)=g(\pm e_{\overline h\alpha})=\pm
  (ge_{\overline h \alpha}) =\pm(\pm e_{\overline g(\overline h
  \alpha)})=\pm e_{(\overline g\overline h)\alpha}=\pm e_{\overline
  {gh}\alpha}. $$
So we only need to prove the equality for some set of generators for
$G_1$.

First consider $\sigma$ and $\rho$, which generate $B$. We have
$$ \sigma(e_\alpha)=\sigma(e_{i_1j_1}\otimes e_{i_2j_2}\otimes
  e_{i_3j_3})=  e_{i_3j_3}\otimes   e_{i_1j_1}\otimes e_{i_2j_2}=
  e_\beta, $$
where $\beta=((i_3,j_3),(i_1,j_1),(i_2,j_2))=\overline\sigma\alpha$,
as $\overline\sigma=(1,(123))$. Similarly
$$ \rho(e_\alpha)=\rho(e_{i_1j_1}\otimes e_{i_2j_2}\otimes e_{i_3j_3})
  = e_{j_2i_2}\otimes e_{j_1i_1}  \otimes e_{j_3i_3}=e_\beta,$$
where $\beta=((j_2,i_2),(j_1,i_1),(j_3,i_3))=(1,(12))\alpha=\overline
\rho\alpha$.

Next consider elements of $A_1$. Any of these elements is $c\widehat\pi$,
where $c\in C_1$ and $\widehat\pi$ is a permutation matrix. An element
of $C_1$ takes any $e_\alpha$ to $\pm e_\alpha$, and $\overline c=1$
($={\rm id}_F$, to be precise). So $ce_\alpha=\pm e_{\overline c\alpha}$
is evident. Next, it is easy to show that for any matrix unity $e_{ij}$
and any permutation $\pi\in S_3$ the equality $\widehat\pi e_{ij}
\widehat\pi^{-1}=e_{\pi i,\pi j}$ is true. Hence for $\alpha=((i_1,j_1),
(i_2,j_2),(i_3,j_3))$ we have
$$ \widehat\pi(e_\alpha)=\widehat\pi e_{i_1j_1}\widehat\pi^{-1}\otimes
  \widehat\pi e_{i_2j_2} \widehat\pi^{-1}\otimes\widehat\pi e_{i_3j_3}
  \widehat\pi^{-1}= e_{\pi i_1,\pi j_1}  \otimes e_{\pi i_2,\pi j_2}
  \otimes e_{\pi i_3,\pi j_3}=e_\beta,$$
where $\beta=((\pi i_1,\pi j_1), (\pi i_2,\pi j_2),(\pi i_3,\pi j_3))
=(\pi,1)\ \alpha=\overline{\widehat\pi}\alpha$. That is, $ge_\alpha=
e_{\overline g\alpha}$ if $g=\widehat\pi$.
\hfill $\square$ \smallskip

We shall call $\alpha=((i_1,j_1),(i_2,j_2),(i_3,j_3))$ {\em even}
if any $m=1,2,3$ occurs evenly many times among $i_1\,,\ldots,j_3$.
For instance, $((1,3),(1,3),(3,3))$ is even and $((1,3),(2,3),(3,1))$
is not. It is clear that the set of even elements of $F$ is invariant
under $S_3\times S_3$.

In the following proposition, and in the sequel, we write ``$11,12,21$''
instead of $((1,1),(1,2),(2,1))$ etc., for brevity.
\smallskip

{\bf Proposition 4.} {\em The group $S_3\times S_3$ has 12 orbits
${\mathcal Q}_1,\ldots,{\mathcal Q}_{12}$ on the set of even elements
of $F$. Their lengths and representatives are listed in the following
table.
\begin{longtable}{|c|c|c||c|c|c||c|c|c|}
\hline
\rule{0pt}{3ex}  $i$ & $\alpha\in{\mathcal Q}_i$ & $|{\mathcal Q}_i|$
  & $i$ & $\alpha\in{\mathcal Q}_i$ & $|{\mathcal Q}_i|$ & $i$
  & $\alpha\in{\mathcal Q}_i$ & $|{\mathcal Q}_i|$ \\
\hline \endhead
\rule{0pt}{3ex}  $1$ & $11,11,11$ & $3$ & $5$ & $11,21,12$ & $18$
   & $9$ & $12,23,31$ & $6$ \\
\hline
\rule{0pt}{3ex}  $2$ & $11,11,22$ & $18$ & $6$ & $11,22,33$ & $6$
  & $10$ & $12,23,13$ & $18$ \\
\hline
\rule{0pt}{3ex}  $3$ & $11,12,21$ & $18$ & $7$ & $11,23,23$ & $18$
   & $11$ & $12,32,13$ & $18$ \\
\hline
\rule{0pt}{3ex}  $4$ & $11,12,12$ & $36$ & $8$ & $11,23,32$ & $18$
   & $12$ & $12,31,23$ & $6$ \\
\hline
\end{longtable}
}

{\bf Proof.} These rather elementary considerations are left to the
reader.
\hfill $\square$ \smallskip

Further we need the following simple lemma.

{\bf Lemma 5.} {\em If $X$ is a linear group acting on a space $V$,
$Y\leq X$ is a subgroup and $v\in V$ is an element such that
$\sum_{y\in Y}yv=0$, then $\sum_{x\in X}xv=0$. }
\smallskip

{\bf Proof.}  Let $g_1,\ldots,g_n$ be the representatives of
cosets $X/Y$. Then
$$\sum_{x\in X}xv= \sum_{i=1}^n \sum_{y\in Y}g_iyv=\sum_{i=1}^n
 g_i(\sum_{y\in Y}yv)=\sum_{i=1}^n g_i(0)=0.$$
\hfill $\square$
\smallskip

{\bf Proposition 6.} {\em 1)  $ce_\alpha=e_\alpha$ for all $c\in C_1$
(or, equivalently, for all $c\in C$) if and only if $\alpha$ is even.

2) If $\alpha$ is even, then $ge_\alpha=e_{\overline g\alpha}$ for any
$g\in G$ (or for any $g\in G_1$). In other words, $G$ permutes
$e_\alpha$, where $\alpha$ is even, always with the plus sign.

3) If $\alpha$ is not even, then $\sum_{g\in G}ge_\alpha=0$.

4) For $1\leq i\leq12$ let $\gamma_i=\sum_{\alpha\in{\mathcal Q}_i}
e_\alpha$. Then the elements $\gamma_i$ constitute a basis of $N^G$.

5) For an element $w\in N$ its projection to $N^G$ is equal to
$$ p(w)={\rm pr}_{N^G}(w)=\sum_{i=1}^{12} (1/|{\mathcal Q}_i|) r_i(w)
  \gamma_i\,,  \eqno (3)$$
where $r_i(w)$ is the sum of coefficients in $w$ at all $e_\alpha$
with $\alpha\in{\mathcal Q}_i$. }
\smallskip

{\bf Proof.} 1) This is easy. For instance, if $\alpha=
i_1j_1,i_2j_2,i_3j_3$ and $c={\rm diag}(-1,1,1)$, then $ce_\alpha=
 (-1)^me_\alpha$, if exactly $m$ of $i_1,j_1,\ldots,j_3$ are equal to
$1$.

2) This easily follows from the arguments in the proof of Lemma 3,
taking into account statement 1), because $\sigma$, $\rho$, and
$\widehat\pi$ permute the tensors $e_\alpha$ (all of them, including
those with $\alpha$ not even) always with plus sign.

3) If $\alpha$ is not even, then by 1) there exists $c\in C$ such that
$ce_\alpha=-e_\alpha$, and we can apply Lemma 5 to the group $X=G$,
subgroup $Y=\{1,c\}$, and the space element $v=e_\alpha$.

4) As the characteristics equals $0$ and $\{e_\alpha\mid \alpha\in F\}$
is a basis of $N$, the elements $\sum_{g\in G}ge_\alpha$ span $N^G$.
If $\alpha$ is not even, then the latter element equals~$0$. If
$\alpha$ is even, this element is a scalar multiple of $\gamma_i$,
where $i$ is such that $\alpha\in{\mathcal Q}_i$. Therefore the elements
$\gamma_i$ span $N^G$. The independence of these elements is obvious.

5) If $\alpha$ is not even, then $G$-average of $e_\alpha$, that is
$p(e_\alpha)$, is $0$. If $\alpha$ is even, then $p(e_\alpha)=
x\gamma_i$, where $i$ is such that $\alpha\in{\mathcal Q}_i$.
The coefficient $x$ can be found using the condition that the sums
of all coefficients, at all $e_\beta$, $\beta\in F$, for $e_\alpha$
and $p(e_\alpha)$ must be the same, whence $1=x|{\mathcal Q}_i|$,
$x=1/|{\mathcal Q}_i|$. Thus, $p(e_\alpha)=(1/|{\mathcal Q}_i|)\gamma_i$.
Hence the formula (3) easily follows.
\hfill $\square$ \medskip

By using the last statement of the proposition we easily can for each
tensor of the form $w_i(a,b,\ldots)$ calculate its orbit sum, i.e.,
the sum of all its $G$-conjugates.

{\bf Example.} Calculate the orbit sum for
\begin{eqnarray*}
	w_{27}(1,2,3,4,5) = && ((e_{11}+e_{22})+2(e_{12}-e_{21})+
3e_{33})\otimes ((e_{11}+e_{22}) \\
   && -2(e_{12}-e_{21})+3e_{33}) \otimes (4(e_{11}+e_{22})+5e_{33}).
\end{eqnarray*}
Make the table containing all even $\alpha$ such that $w$ involves
$e_\alpha$, with the corresponding coefficient $v_\alpha$ and the
number $1\leq i\leq12$ such that $\alpha\in{\mathcal Q}_i$.
\begin{longtable}{|c|c|c||c|c|c||c|c|c|}
\hline
\rule{0pt}{3ex}  $\alpha$ & $i$ & $v_\alpha$ & $\alpha$ & $i$
  & $v_\alpha$ & $\alpha$ & $i$ & $v_\alpha$ \\
\hline \endhead
\rule{0pt}{3ex} $11,11,11$ & $1$ & $4$ & $11,11,22$ & $2$ & $4$
  & $11,11,33$ & $2$ & $5$  \\
\hline
\rule{0pt}{3ex} $11,22,11$ & $2$ & $4$ & $11,22,22$ & $2$ & $4$
   & $11,22,33$ & $6$ & $5$  \\
\hline
\rule{0pt}{3ex} $22,11,11$ & $2$ & $4$ & $22,11,22$ & $2$ & $4$
   & $22,11,33$ & $6$ & $5$  \\
\hline
\rule{0pt}{3ex} $22,22,11$ & $2$ & $4$ & $22,22,22$ & $1$ & $4$
  & $22,22,33$ & $2$ & $5$  \\
\hline
\rule{0pt}{3ex} $11,33,11$ & $2$ & $12$ & $11,33,22$ & $6$ & $12$
   & $11,33,33$ & $2$ & $15$  \\
\hline
\rule{0pt}{3ex} $22,33,11$ & $6$ & $12$ & $22,33,22$ & $2$ & $12$
   & $22,33,33$ & $2$ & $15$  \\
\hline
\rule{0pt}{3ex} $33,11,11$ & $2$ & $12$ & $33,11,22$ & $6$ & $12$
   & $33,11,33$ & $2$ & $15$  \\
\hline
\rule{0pt}{3ex} $33,22,11$ & $6$ & $12$ & $33,22,22$ & $2$ & $12$
   & $33,22,33$ & $2$ & $15$  \\
\hline
\rule{0pt}{3ex} $33,33,11$ & $2$ & $36$ & $33,33,22$ & $2$ & $36$
   & $33,33,33$ & $1$ & $45$  \\
\hline
\rule{0pt}{3ex} $12,12,11$ & $4$ & $-16$ & $12,12,22$ & $4$ & $-16$
   & $12,12,33$ & $7$ & $-20$  \\
\hline
\rule{0pt}{3ex} $12,21,11$ & $3$ & $16$ & $12,21,22$ & $5$ & $16$
  & $12,21,33$ & $8$ & $20$  \\
\hline
\rule{0pt}{3ex} $21,12,11$ & $5$ & $16$ & $21,12,22$ & $3$ & $16$
  & $21,12,33$ & $8$ & $20$  \\
\hline
\rule{0pt}{3ex} $21,21,11$ & $4$ & $-16$ & $21,21,22$ & $4$ & $-16$
  & $21,21,33$ & $7$ & $-20$  \\
\hline
\end{longtable}

Using this table we can find the coefficients of the orbit sum in
the basis $\{\gamma_i\}$. As an example, find the coefficient at
$\gamma_1$. The coefficients in $w$ at $e_{11,11,11}=e_{11}\otimes
e_{11}\otimes e_{11}$, $e_{22,22,22}$, and $e_{33,33,33}$ are $4$,
$4$, and $45$, respectively. The coefficient in $p(w)$ at
$\gamma_1$ is $r_1(w)/|{\mathcal Q}_1|=(4+4+45)/3=53/3$, according
to Proposition 6.5). The orbit of $w$ has length 18, whence the
orbit sum is $18p(w)$, and the coefficient at $\gamma_1$ in this
sum is $53\cdot6=318$. Similarly one can calculate the other
coefficients (which is recommended to the reader as an exercise)
and find the complete orbit sum, which is equal to
$$ 318\gamma_1+214\gamma_2+32\gamma_3-32\gamma_4+32\gamma_5+174
  \gamma_6-40\gamma_7+40\gamma_8 $$
(note that $\gamma_9,\ldots,\gamma_{12}$ are not involved in this sum).

Thus, we see that the calculation turns out to be rather long. However, to prove Theorem 1
we shall not need the orbit sums for all tensors $w_i(a,b,\ldots)$
for arbitrary $a,b,\ldots$! Knowing the coefficients at {\em some}
$\gamma_i$ in {\em some} sums will be sufficient.
\medskip

{\bf 4. The proof of Theorem 1.} Now we can start proving Theorem~1.
Assume on the contrary that a $G$-invariant decomposition of length
$\leq23$ for ${\mathcal T}$ does exist, and among all such
decompositions take the one of the smallest length.
\smallskip

{\bf Proposition 7.} {\em 1) A minimal $G$-invariant decomposition
for ${\mathcal T}$ does not contain an orbit of any of the types $16$,
$18$, $21$, $25$, $33$, or $42$.

2) There exists a minimal decomposition not containing orbits of type
$4$, $39$, or $43$. }
\smallskip

{\bf Proof.} 1) Consider three tensors $w'=w_7(1)=\delta^{\otimes3}$,
$w''=w_6(1)=\eta^{\otimes3}$, and $w'''=w_5(0,1)=e_{33}^{\otimes3}$.
Their orbits are ${\mathcal O}'=\{\delta^{\otimes3}\}$,
${\mathcal O}''=\{\eta^{\otimes3},\overline\eta^{\otimes3}\}$, and
${\mathcal O}'''=\{e_{11}^{\otimes3},e_{22}^{\otimes3},
e_{33}^{\otimes3}\}$, respectively, and the orbit sums are
$\sigma'=\gamma_1+\gamma_2+\gamma_6$, $\sigma''=2\gamma_1-\gamma_2
+2\gamma_6$, and $\sigma'''=\gamma_1$. So any linear combination of
$\gamma_1$, $\gamma_2$, and $\gamma_6$ is a linear combination of
$\sigma'$, $\sigma''$, and $\sigma'''$ and can be therefore expressed
as a sum of some $G$-invariant set of decomposable tensors of
$\leq6$ elements.

Note that for any $i\in\{16,18,21,25,33,42\}$ the tensor
$w_i(a,b,\ldots)$ involves summands of the forms $e_{jj}\otimes e_{kk}
\otimes e_{ll}$ only, so its orbit sum is a linear combination of
$\gamma_1$, $\gamma_2$, and $\gamma_6$. Therefore, this orbit sum is
a sum of a $G$-invariant set of decomposable tensors of $\leq6$
elements. But this orbit contains $>6$ tensors. Thus it can be replaced
by a {\em smaller} $G$-invariant set of decomposable tensors with
the same sum. This contradicts the assumption that the decomposition
under consideration is of minimal possible length.

2) The argument is similar. An orbit of each of the types $4$, $39$,
or $43$ can be replaced by a union of orbits of types $5$, $6$, and
$7$ having the same sum. Since the length of an orbit of type $4$,
$39$, or $43$ is $6$, the overall length of the decomposition does
not increase after such a replacement.   \hfill $\square$
\smallskip

{\bf Lemma 8.} {\em The orbit sum for the tensor $w_9(a,b)=(a\delta+
b\varkappa)^{\otimes3}$ is $4b^3(\gamma_9+\gamma_{10}+\gamma_{11}+
\gamma_{12})+D$, where $D\in\langle \gamma_1,\ldots,\gamma_8\rangle$. }
\smallskip

{\bf Proof.} We have $N=N_1\oplus N_2$, where $N_1$ is the span of
all $e_\alpha$ such that $\alpha\in {\mathcal Q}_i$, $i=9,10,11,12$,
i.e., of
all $e_{i_1j_1}\otimes e_{i_2j_2}\otimes e_{i_3j_3}$ such that
$\{ \{i_1,j_1\}, \{i_2,j_2\}, \{i_3,j_3\} \}= \{ \{1,2\}, \{2,3\},
\{1,3\} \}$, and $N_2$ is the span of remaining $e_\alpha$. It is
clear that both $N_1$ and $N_2$ are $G$-invariant. For a tensor
$t\in N$ let $t_1$ and $t_2$ be its $N_1$- and $N_2$-components.

It is more or less obvious that $[(a\delta+b\varkappa)^{\otimes3}]_1=
[(b\varkappa)^{\otimes3}]_1=b^3[\varkappa^{\otimes3}]_1$. Next, it is
clear that $[\varkappa^{\otimes3}]_1$ is the sum of all $e_\alpha$ such
that $\alpha\in{\mathcal Q}_i$, $i=9,10,11,12$, and the latter sum
is, clearly,
nothing else but $\gamma_9+\gamma_{10}+\gamma_{11}+\gamma_{12}$. Thus,
$[w_9(a,b)]_1=b^3(\gamma_9+\gamma_{10}+\gamma_{11}+\gamma_{12})$.
So the orbit sum for $w_9(a,b)$ is
\begin{eqnarray*} 4p(w_9(a,b)) &=& 4p((w_9(a,b))_1+(w_9(a,b))_2)
  =D+4p((w_9(a,b))_1) \\
 &=& D+4p(b^3(\gamma_9+\gamma_{10}+\gamma_{11}+\gamma_{12}))=
4b^3(\gamma_9+\gamma_{10}+\gamma_{11}+\gamma_{12})+D,
\end{eqnarray*}
where $D=4p((w_9(a,b))_2)$. Finally, it is clear that $p(x)\in\langle
\gamma_1,\ldots,\gamma_8\rangle$ for any $x\in N_2$.
\hfill $\square$ \smallskip

{\bf Proposition 9.} {\em A $G$-invariant decomposition of length
$\leq23$ can not contain an orbit of any of the types $17,22,23,26,
27,28,30,31,36,37$. }
\smallskip

{\bf Proof.}  Let $I=\{17,22,23,26,27,28,30,31,36,37\}$ be the set of
types listed in the hypothesis. Assume on the contrary that a
decomposition containing an orbit ${\mathcal O}$ of a type $i\in I$
does exist. Since an orbit of any type $i\in I$ is of length $18$, the
rest of the decomposition contains $\leq5$ tensors, and so can only
contain orbits of types $5$, $6$, $7$, or $9$.

We can immediately see from the table of orbits that the tensor
$w_i(a,b,\ldots)$ with $i\in I$ does not involve summands proportional
to $e_\alpha$, $\alpha\in{\mathcal Q}_j$, $j=9,10,11,12$. Therefore
its orbit sum does not involve such summands also, and so is in
$\langle\gamma_1,\ldots,\gamma_8\rangle$. The same is true for
$i=5,6,7$. But ${\mathcal T}=\gamma_1+\gamma_3+\gamma_9$. So the
decomposition necessary contains an orbit of type $9$, that is, the
orbit of the tensor $w_9(a,b)=(a\delta+b\varkappa)^{\otimes3}$ with
$b\ne0$. By Lemma 8 the orbit sum of the latter tensor is
$4b^3(\gamma_9+ \gamma_{10}+ \gamma_{11}+ \gamma_{12})+D$, where
$D\in\langle \gamma_1,\ldots,\gamma_8\rangle$. So the sum of all the
tensors of the decomposition involves $\gamma_9$, $\gamma_{10}$,
$\gamma_{11}$, and $\gamma_{12}$ with the same coefficients --- but
this is not the case for ${\mathcal T}$.
\hfill $\square$ \medskip

Our next aim is to eliminate the remaining orbits of length $18$.
\smallskip

{\bf Lemma 10.} {\em Let $w=w_l(a,b,\ldots)$ be a decomposable tensor
of type $l=24,29,32,38$, and $s$ be its orbit sum. Then the coefficients
in $s$ at $\gamma_m$, where $m=9,\ldots,12$, are listed in the following
table:

\begin{longtable}{|c|c|c|c|c|}
\hline
\rule{0pt}{3ex}  & $24$ & $29$ & $32$ & $38$ \\
\hline \endhead
\rule{0pt}{3ex} $\gamma_9$ & $6a^2d$ & $6ia^2d$ & $6a^2d$ & $6ia^2d$ \\
\hline
\rule{0pt}{3ex} $\gamma_{10}$ & $2a^2d+4abd$ & $2ia^2d+4iabd$
  & $2a^2d+4abd$ & $-2ia^2d+4iabd$ \\
\hline
\rule{0pt}{3ex} $\gamma_{11}$ & $2b^2d+4abd$ & $2ib^2d+4iabd$
  & $2b^2d-4abd$ & $2ib^2d-4iabd$ \\
\hline
\rule{0pt}{3ex} $\gamma_{12}$ & $6b^2d$ & $6ib^2d$ & $-6b^2d$
  & $-6ib^2d$ \\
\hline
\end{longtable}
}
\smallskip

{\bf Proof.} A direct computation similar to the Example in the end
of Section~3.
\hfill $\square$ \smallskip

{\bf Proposition 11.} {\em A $G$-invariant decomposition for
${\mathcal T}$ of length $\leq23$ can not contain an orbit of any
of types $l=24,29,32,38$. }
\smallskip

{\bf Proof.}  Assume on the contrary that such a decomposition does
exist. Then ${\mathcal T}=s_l+s'$, where $s_l$ is the orbit sum for
$w_l(a,b,\ldots)$, containing $18$ summands, and $s'$ is the sum
of the remaining
summands. Obviously, $s'$ contains $\leq5$ summands (tensors).
So one of the following cases holds: (a) $s'$ contains an orbit of
length $4$ (and therefore of type $9$), and may be an orbit of type
$7$, that is, a multiple of $\delta^{\otimes3}$, or (b) $s'$
only contains orbits of types $5$, $6$, or $7$. We take these two
cases to a contradiction separately.

(a) In this case $s'$ is the sum of two summands, namely the orbit
sum for $w_9(a,b)=(a\delta+b\varkappa)^{\otimes3}$ and another
summand $c\delta^{\otimes3}$. Note that
$w_l$ and therefore $s_l$ does not involve any summands proportional to
$e_{ii,jj,kk}$. On the other hand, in $(a\delta+b\varkappa)^{\otimes3}$
such summands are the same as in $(a\delta)^{\otimes3}$, with the same
coefficients. Therefore the sum of all summands of this form
in ${\mathcal T}=s_l+s'$ is the same as in $(c+4a^3)\delta^{\otimes3}$.
But this contradicts to the fact that ${\mathcal T}$ involves
$e_{11,11,11}$ but not $e_{11,11,22}$.

(b) In this case, obviously, $s'$ does not involve $\gamma_m$ with
$m=9,\ldots,12$. Since ${\mathcal T}$ involves $\gamma_9$, but not
$\gamma_{10}$, $\gamma_{11}$, or $\gamma_{12}$, we conclude that
$s_l$ also involves $\gamma_9$, but not $\gamma_{10,11,12}$. By
Lemma 10 the condition that ${\mathcal T}$ involves $\gamma_9$ implies
$a^2d\ne0$, and the condition that ${\mathcal T}$ does not involve
$\gamma_{12}$ implies $b^2d=0$. Then $a,d\ne0$ and $b=0$, whence
the coefficient in ${\mathcal T}$ at $\gamma_{10}$ is not equal to
$0$, a contradiction.
\hfill $\square$ \smallskip

{\bf Proposition 12.} {\em A $G$-invariant decomposition for ${\mathcal T}$
of length $\leq23$ does not contain an orbit of type $35$. }
\smallskip

{\bf Proof.} In the same way like in the previous Proposition we have
two cases (a) and (b). In the case (a) the contradiction can be
obtained by the same argument. As to (b) case, note that neither
the orbit sum for $w_{35}(a,b,\ldots)$ nor $s'$ can involve a summand
proportional to $e_{11,12,21}$. But ${\mathcal T}$ involves such
a summand.
\hfill $\square$ \smallskip

{\bf Lemma 13.} {\em For any tensor $w=u\otimes u\otimes v$ the sum
$s=\sum_{g\in G} gw$ involves $\gamma_3$ and $\gamma_5$ with the same
coefficients. }
\smallskip

{\bf Proof.}  Let $\pi_{12}\colon x\otimes y\otimes z\mapsto y\otimes
x\otimes z$ be the usual (i.e., without transposing of matrices)
transposition of the first two factors in the tensor cube $M\otimes M
\otimes M$. Obviously,
$\pi_{12}w=w$. Clearly, $\pi_{12}$ commutes with any element $a\in A$.
It is also easy to see that $\pi_{12}$ commutes with $\rho\in B$,
and the conjugation by $\pi_{12}$ inverts $\sigma$. So $\pi_{12}$
normalizes $G$, $\pi_{12}G\pi_{12}=G$, $\pi_{12}G=G\pi_{12}$. Now
we have
$$ \pi_{12}s = \pi_{12}(\sum_{g\in G} gw)=\sum_{g\in G} (\pi_{12}g)w
  =\sum_{g\in G}(g\pi_{12})w  = \sum_{g\in G} g(\pi_{12}w)=
  \sum_{g\in G} gw=s. $$
Further, observe that $\pi_{12}$ preserves the set of all tensors
$e_\alpha$ and leaves the set of all $e_\alpha$ with $\alpha$ even
invariant. Since $\pi_{12}$ normalizes $G$, it preserves the partition
of the set $\{e_\alpha\}$ with even $\alpha$ into $G$-orbits,
and therefore permutes $\{\gamma_i\mid i=1,\ldots,12\}$.

It is clear that $\pi_{12}$ permutes $e_{12,21,11}$ with $e_{21,12,11}$.
So it permutes the orbit sum for $e_{12,21,11}$, which is equal to
$\gamma_3$, with the orbit sum for $e_{21,12,11}$ which is equal to
$\gamma_5$.

If $s=a\gamma_3+b\gamma_5+z$, where $z\in L:=\langle\gamma_i\mid i
\ne 3,5\rangle$, then $s=\pi_{12}s=a\gamma_5+b\gamma_3+z'$, where
$z'\in L$ also. So $a=b$.
\hfill $\square$ \smallskip

Now we can finish the proof of Theorem~1. Assume on the contrary that
there exists a $G$-invariant decomposition ${\mathcal P}$ for
${\mathcal T}$ of length $\leq23$.
By Proposition 7.2) we can assume that ${\mathcal P}$ contains
no orbits of type $4$, $39$, or $43$. Next, ${\mathcal P}$ contains
no orbits of types $16,18,21,25,33,42$ by Proposition 7.1); orbits
of types $17,22,23,26,27,28,30,31,36$, or $37$ by Proposition 9;
of types $24,29,32,38$ by Proposition 11; and orbits of type $35$ by
Proposition 12. The remaining types are the following: $1,\ldots,15$,
except for $4$; and $19,20,34,40,41,44$. For each of these
types, except for $44$, the tensor $w_i(a,b)$ is of the form
$u^{\otimes2}\otimes v$, and therefore its orbit sum involves
$\gamma_3$ and $\gamma_5$ with the same coefficients. Also, for type
$44$ the orbit sum does not involve neither $\gamma_3$ nor $\gamma_5$,
because $w_{44}$ does not involve $e_\alpha$ such that
$\alpha\in{\mathcal Q}_3$ or $\alpha\in{\mathcal Q}_5$. Therefore,
${\mathcal T}$ must involve $\gamma_3$ and $\gamma_5$ with the same
coefficients, a contradiction.

The proof of Theorem 1 is complete.

\smallskip

\begin{center}\textbf{References}\end{center}

1. Burichenko V.P., Non-existence of a short algorithm for multiplication 
of $3\times 3$ matrices with group $S_4\times S_3$
// Труды Института математики (=Proceedings of the Institute of
mathematics), accepted for publication. See also arXiv: 2211.03404, 2022. 

2. Burichenko V.P., Symmetries of matrix multiplication algorithms. I // 
 arXiv: 1508.01110, 2015. 

3. Burichenko V.P., The isotropy group of the matrix multiplication tensor // 
Труды Института математики (=Proceedings of the Institute of
mathematics), 24:2 (2016), 106--118. See also arXiv: 2210.16565, 2022. 

4. Brent R.P., Algorithms for matrix multiplication // Technical report 70-157,
Stanford university, Computer Science Department, 1970. 
\\ Available at {\tt http://maths-people.anu.edu.au/~brent/pub/pub002.html}.

\end{document}